\documentclass[epsfig,graphics,twocolumn,showpacs,floatfix,mathbbm,prl,superscriptaddress]{revtex4}

\usepackage{amsmath,amssymb,graphicx,color}
\usepackage{float}
\usepackage{subfigure}
\usepackage{amsmath}
\newcommand{\PrYSO}{Pr$^{3+}$:Y$_2$Si{O$_5$}}
\newcommand{\YSO}{Y$_2$Si{O$_5$}}

 \newcommand{\F}[1]{\mathcal{F}_{{#1}}}
 \newcommand{\SNR}{\textrm{SNR}}

\begin{document}

\title{A solid state spin-wave quantum memory for time-bin qubits}

\author{Mustafa G\"{u}ndo\u{g}an}
\author{Patrick M. Ledingham}
\altaffiliation{Present address: Department of Physics, Clarendon Laboratory, University of Oxford, Oxford OX1 3PU, United Kingdom}
\author{Kutlu Kutluer}
\author{Margherita Mazzera}
\email{margherita.mazzera@icfo.es}
\affiliation{ICFO-Institut de Ciencies Fotoniques, Mediterranean Technology Park, 08860 Castelldefels (Barcelona), Spain}
\author{Hugues de Riedmatten}
\affiliation{ICFO-Institut de Ciencies Fotoniques, Mediterranean Technology Park, 08860 Castelldefels (Barcelona), Spain}
\affiliation{ICREA-Instituci\'{o} Catalana de Recerca i Estudis Avan\c cats, 08015 Barcelona, Spain}

\date{\today}
\begin{abstract} 
We demonstrate the first solid-state spin-wave optical quantum memory with on-demand read-out. Using the full atomic frequency comb scheme in a \PrYSO crystal, we store weak coherent pulses at the single-photon level with a signal to noise ratio $> 10$.  Narrow-band spectral filtering based on spectral hole burning in a second \PrYSO crystal is used to filter out the excess noise created by control pulses to reach an unconditional noise level of  $(2.0 \pm 0.3) \times10^{-3}$ photons per pulse. We also report spin-wave storage of photonic time-bin qubits with conditional fidelities higher than a measure and prepare strategy, demonstrating that the spin-wave memory operates in the quantum regime. This makes our device the first demonstration of a quantum memory for time-bin qubits, with on demand read-out of the stored quantum information. These results represent an important step for the use of solid-state quantum memories in scalable quantum networks.
\end{abstract}

\maketitle

Photonic quantum memories are essential in
quantum information science (QIS) where they are used as quantum
interfaces between flying and stationary qubits. They enable the
synchronization of probabilistic quantum processes e.g. in quantum
communication \cite{Duan2001,Sangouard2011} and computing \cite{Knill2001}.
The implementation of quantum memories (QMs) for light requires
strong interactions between individual photons and matter. This
can be achieved by placing individual quantum systems (e.g. single
atoms) in high finesse cavities \cite{Specht2011} or by using
ensembles of atoms, where the photons are mapped onto collective
atomic excitations. Atomic systems are natural candidates as QMs
\cite{Chou2005,Chaneliere2005,Choi2008,Zhang2011,Lettner2011,Riedl2012,Nicolas2014,Julsgaard2004,Hosseini2011,Sprague2014},
but solid state systems offer interesting perspectives for
scalability and integration into existing technology \cite{Riedmatten2008,Hedges2010,Clausen2011,Saglamyurek2011,Togan2010,Bernien2013,England2015}.

Rare-earth ion doped solids are promising candidates for high
performance solid state QMs since they have excellent coherence
properties at cryogenic temperatures \cite{Macfarlane2002}. They
also exhibit large static inhomogeneous broadening of the optical
transitions which can be tailored and used as a resource for
various storage protocols, e.g. enabling temporally
\cite{Afzelius2009} and spectrally \cite{Sinclair2014} multiplexed
quantum memories. Recent experimental progress includes qubit storage \cite{Riedmatten2008,Gundogan2012,Clausen2012,Zhou2012, Sinclair2014}, highly
efficient quantum storage of weak coherent states
\cite{Hedges2010}, storage of entangled and single photons
\cite{Clausen2011,Saglamyurek2011,Rielander2014}, entanglement
between two crystals \cite{Usmani2012} and quantum teleportation
\cite{Bussieres2014}.

Yet, nonclassical states have so far only been stored as
collective optical atomic excitations with fixed storage times
\cite{Clausen2011,Saglamyurek2011,Rielander2014}. While this may
provide a useful resource if combined with massive multiplexing
and deterministic quantum light sources \cite{Sinclair2014}, the
ability to read-out the stored state on-demand is essential for
applications where the quantum memory is used as a synchronizing
device. On-demand read-out can be achieved by actively controlling
the optical collective excitations \cite{Hedges2010}, with a
storage time limited by the coherence of excited states. Another
solution is to transfer the optical excitations to long lived
collective spin excitations (or spin-waves), using strong control
pulses \cite{Afzelius2010,Gundogan2013,Timoney2013}. This gives access to much longer storage times
\cite{Longdell2005, Heinze2013}. Operating a solid state spin-wave
memory in the quantum regime has so far remained elusive, because
of insufficient signal-to-noise ratio (SNR) at the single-photon level \cite{Timoney2013}.

Here, using the full atomic frequency comb (AFC) protocol in a
\PrYSO crystal we store and retrieve weak coherent pulses
on-demand with SNR $>$ 10 for single-photon-level input. Using a
narrowband filter based on spectral hole burning in a second
crystal, we achieve an unconditional noise floor of $(2.0 \pm 0.3)
\times10^{-3}$ photons per pulse. The use of spectral holes as narrowband filters has been already demonstrated in storage schemes operating in a classical fashion \cite{Zhang2012,Beavan2012,Beavan2013}, but it is here exploited for the first time to enter the quantum regime. Finally, we demonstrate storage
and retrieval conditional fidelities (i.e., assuming that a
photon was reemitted)  higher than  classical memories for
time-bin qubits at the single-photon level, taking into account the Poissonian statistics and the finite efficiency of the memory. These results
represent the first demonstration of a solid state spin-wave
quantum memory, enabling on-demand read-out of the stored qubits.
They also provide the first example of a spin-wave quantum memory
for time-bin qubits (for any system), an essential resource in
quantum communication \cite{Marcikic2003} and processing
\cite{Humphreys2013}.

The AFC technique \cite{Afzelius2009,Riedmatten2008} is based on
spectral tailoring of an inhomogeously broadened absorption line
into a comb-shaped structure with periodicity $\Delta$. The input
pulses resonant with the comb are mapped onto a collective optical
atomic excitation. After an initial dephasing, the excitations
rephase at a time $1/\Delta$ giving rise to a forward collective
emission \cite{Afzelius2009,Riedmatten2008}. Before the coherent
emission two strong control pulses are applied to transfer the
excitation to and from a long-lived ground state to achieve the
spin-wave storage of the input pulses. The full AFC scheme
requires ions with at least three ground states, one being used as auxiliary state for optical pumping
\cite{SupplMat}. The spin-wave
storage efficiency is given by $\eta_{SW} = \eta_{AFC} \times
\eta_T^2 \times \eta_C$, where $\eta_{AFC}$ is the efficiency of
the storage at the excited state and depends on the optical depth
and comb finesse \cite{Afzelius2009}, $\eta_C$ accounts for the
decoherence during the ground state storage, and $\eta_T$ is the
transfer efficiency of the control pulses.

\begin{figure}[htbp]
\centerline{\includegraphics[width=1\columnwidth]{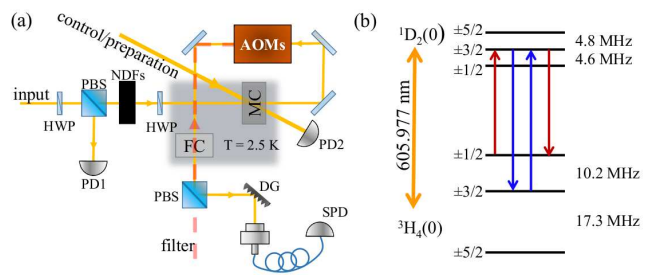}}
\caption{(a) \textbf{Quantum memory setup}. The memory (MC) and
filter (FC) crystals are located inside a liquid-free cooler
(Oxford V14) operating at a temperature of $2.5\,\mathrm{K}$. They are both $3\,\mathrm{mm}$ long and doped with a Pr$^{3+}$ concentration of $0.05 \,\%$. 
The control and input beams are steered towards the memory with an angle of $\sim1.5\,\mathrm{^{\circ}}$, leading to a exctinction ratio of 10$^{-5}$. 
The beam diameters on the crystal are $280\,\mathrm{\mu m}$ and $90\,\mathrm{\mu m}$ for strong and input modes, respectively. The
weak coherent states are prepared by attenuating bright pulses
with variable neutral density filters (NDFs). A portion of the
input beam is picked up before the NDF and sent to a photodiode
(PD1) for the calibration of the mean photon number per pulse.  A mechanical shutter is used to protect the SPD during the memory and filter preparation.
HWP: half-wave plate; AOM: acousto-optical modulator; DG: diffraction grating; SPD: single-photon detector. 
The dashed red beam indicates the filter preparation mode.
(b) Hyperfine splitting of the first sublevels of the ground
$^{3}$H$_{4}$ and the excited $^{1}$D$_{2}$ manifold of Pr$^{3+}$ in \YSO.
} \label{setup}
\end{figure}

The realization of the full AFC scheme in the single-photon regime
is challenging as the strong control pulses create noise which
may dominate the weak signal retrieved from the memory. Two main
mechanisms are responsible for this noise. i) Spatial leakage from
the control mode into the input mode due to scattering from the
optical surfaces and ii) interaction of the control pulses with
residual population in the spin storage state due to
imperfect optical pumping. The latter includes collective effects,
such as free-induction decay (FID), or incoherent fluorescent emission.
Note that  four-wave mixing is not a dominant source of noise in our system, in contrast to Raman memories in atomic vapors \cite{Michelberger2015} (see \cite{SupplMat} for details).
To reduce the noise  we employ spatial, temporal and spectral
filtering.
The spectral filtering is challenging in \PrYSO, as the input and
control frequencies are separated by only $ 10.2\,\mathrm{MHz}$ (see Fig.
\ref{setup}(b)). As a narrow-band spectral filter we use a second \PrYSO crystal where we prepare spectral holes of variable width \cite{Zhang2012,Beavan2012,Beavan2013}.

The experimental arrangement and the relevant energy level scheme  of {Pr$^{3+}$} at
$606\,\mathrm{nm}$ are shown in Fig. \ref{setup}. 
The main laser beam at $606\,\mathrm{nm}$
(Toptica TA-SHG pro) is split into three to be used as input mode,
filter preparation, and lastly for control pulses and memory
preparation. They all pass through acousto-optical modulators
(AOMs) in double-pass configuration, driven by an arbitrary
waveform generator (Signadyne), to create the necessary pulse
sequences. The beams are then carried with
polarization-maintaining single-mode optical fibers to another
optical table where the cryostat is located. The maximum available
optical powers are about 20 mW, 3.5 mW and 150 $\mu$W for control,
filter preparation, and input modes, respectively, measured in
front of the cryostat. The frequency of the $606\,\mathrm{nm}$
laser is stabilized by Pound-Drever-Hall technique to a home-made
temperature controlled Fabry-Perot cavity housed in a vacuum
chamber. The input light is linearly polarized close to the
optical $D_2$ axis to maximize the interaction with the Pr$^{3+}$
ions. The measured optical depth of the Pr$^{3+}$ transition at $606\,\mathrm{nm}$ is about 7 for both memory and filter crystal. In both cases, the inhomogeneous linewidth is about $6\, \mathrm{GHz}$. After the storage, the retrieved signal passes through different diffraction order modes 
(-1st and 1st) of two consecutive AOMs, acting as temporal gate before passing through the filter crystal.
 A diffraction grating (DG) is then used to filter the noise not resonant
with the crystal. The retrieved signal is coupled with $60 \,\%$
efficiency into a single-mode fiber for connection to the single photon detector (SPD, PicoQuant $\tau$ SPAD-20, detection efficiency $\eta_d = 60\,\%$, dark
count rate $\sim 10 \,\mathrm{Hz}$).  The total transmission of the input beam
from the cryostat to the SPD is about $13 \,\%$.

We tailor the AFC using  optical pumping techniques as described in \cite{Timoney2013,Maring2014}. We isolate a single class of atoms \cite{Nilsson2004,Guillot-Noel2009} and create
a $ 3.5 \,\mathrm{MHz}$ wide AFC with $\Delta$ = 200 kHz on the
${1}/{2}_\textrm{g}-{3}/{2}_\textrm{e}$ transition within a 14 MHz
wide transparency window (see \cite{SupplMat} for details and for a comb example). 
During the preparation of the memory,
the population removed from the comb is stored in the auxiliary
${5}/{2}_\textrm{g}$ state, while the ${3}/{2}_\textrm{g}$ is
emptied. To further remove unwanted residual population in the ${3}/{2}_\textrm{g}$ state,   we apply an extra series of 100  pulses on the ${3}/{2}_\textrm{e}-{3}/{2}_\textrm{g}$ transition after
the comb preparation. We then start the single-photon-level
storage measurements. 
Weak  gaussian input  pulses with full width at half maximum (FWHM) duration of 430 ns and  mean photon number $\mu_{in}$ are mapped on the AFC and transferred to spin-waves
thanks to a strong control pulse. 
The control pulses have a Gaussian temporal profile
with FWHM of 700 ns and are spectrally
chirped by $5\,\mathrm{MHz}$ about the ${3}/{2}_\textrm{e}-{3}/{2}_\textrm{g}$ transition. For each comb preparation, 1000 storage trials are perfomed  with a repetition rate of $\sim
7 \,\mathrm{kHz}$. The full cycle has a period of 700 ms,  including  memory preparation and light storage. It is 
synchronized with the cryostat cycle to reduce the effect of vibrations. 
The sequence is then repeated 500 to 1000 times to acumulate sufficient statistics.  
 
\begin{figure}
\centerline{\includegraphics[width=1\columnwidth]{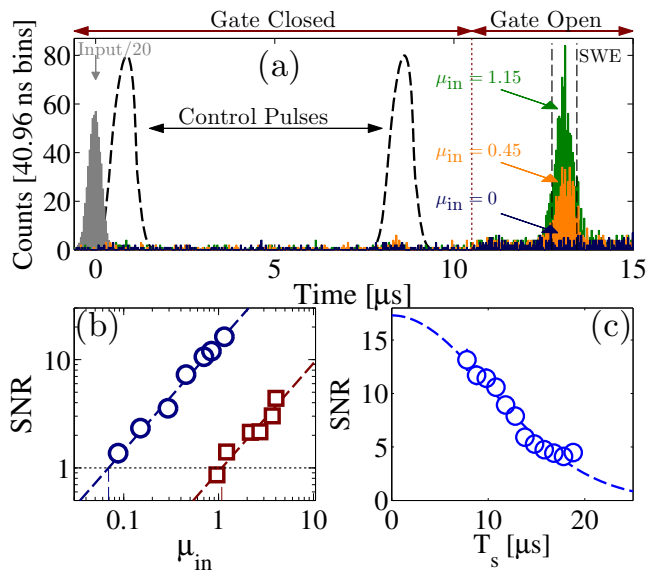}}
\caption{ (a). Time histograms of the retrieved photons measured
for different input photon numbers when a transparency window $
2\,\mathrm{MHz}$ wide is prepared in the filter crystal. The input
($\mu_{in} = 0.9$) and the control pulses, as measured in photon
counting and from a reference photodiode (PD2 in Fig.
\ref{setup}), respectively, are also displayed. The chosen
$0.7\,\mathrm{\mu s}$ wide detection window is indicated by the
dashed lines about the three-level echo; it includes $\sim 80\,\%$
of the counts in the full echo mode. (b). Signal-to-noise ratio
(SNR) as a function of the number of input photons for different
filter width. Circles: $ 2\,\mathrm{MHz}$; Squares: $
14\,\mathrm{MHz}$. The error bars (smaller than the data points)
are evaluated with Poissonian statistics. The black dotted line
indicates the limit of detection SNR = 1. The dashed lines are
linear fits of the experimental data. Panel (c). Decay of the SNR
as a function of the spin-wave storage time T$_S$ with average
photon number $\mu_{in} =1$. 
From the fit with a Gaussian profile, the spin inhomogeneous broadening $\gamma_{in} = (26 \pm 1)
\,\mathrm{kHz}$ can be extrapolated.  By comparing the SNR that we
measure at a storage time $T_S = 7.8\,\mathrm{\mu s }$ with the
extrapolation at $T_S = 0\,\mathrm{\mu s }$, we can evaluate the
decoherence term $\eta_C$ to be about $ 75\,\%$.} 
\label{SNR}
\end{figure}

Fig. \ref{SNR}(a) shows the time histograms of the retrieved photons with different  $\mu_{in}$. For this measurement, the crystal filter has a hole width of 2 MHz. 
The spin-wave storage time is T$_{S}$ = 7.8 $\mu$s, leading to a total storage time of
$\tau_s=1/\Delta+T_{S}=12.8 \mu s$.  From the trace with
$\mu_{in} = 0$, we estimate an unconditional noise floor of $(2.0 \pm 0.3)
\times10^{-3}$ photons per pulse at the memory crystal.
For $\mu_{in} = 1.15$, we measure SNR = $16.3 \pm 2.4$. The
linear scaling of the echo SNR with respect to increasing
$\mu_{in}$ is shown with blue circles in Fig. \ref{SNR}(b).
Typical values of efficiencies are $\eta_{AFC}=(5.6 \pm 0.3)\,\%$ and
$\eta_{SW} = (2.8\pm 0.1) \,\%$, from which we deduce
$\eta_T=(81.7 \pm 2.6)\,\%$ (assuming $\eta_C \sim 75\,\%$ \cite{Gundogan2013}). A convenient figure of merit taking into account the
noise and efficiency is given by the minimum $\mu_{in}$ necessary
to detect a spin-wave echo with SNR = 1, called $\mu_1$. From the
linear fit of the experimental data of Fig. \ref{SNR}(b), we find
$\mu_1 = 0.069 \pm 0.002$. We then vary $\tau_s$ by changing
$T_S$ with $\mu_{in} =1 $, as shown in Fig. \ref{SNR}(c). The decay
in the SNR is compatible with a spin inhomogeneous broadening of
$\gamma_{in} = (26 \pm 1) \,\mathrm{kHz}$, similar to previous
measurements with bright pulses \cite{Afzelius2010, Gundogan2013}.
We still observe SNR $=4.5 \pm 0.4$ for  $T_S= 18.8\,\mathrm{\mu
s}$ ($\tau_s=23.8\,\mathrm{\mu s}$).

We can estimate the contribution of the filter crystal in the
suppression of the noise by preparing a wider transparency window.
For a filter width of $ 14\,\mathrm{MHz}$ (squares in Fig. \ref{SNR}(b))
there is no filtering at the control frequency, and we observe an
increase of the noise floor to $(2.3 \pm 0.6) \times 10^{-2}$
together with a slightly higher retrieval efficiency ($\eta_{SW} =
(2.9 \pm 0.2)\,\%$), which results in $\mu_1$ values up to about
1. When the filter crystal is by-passed, the noise floor raises to
$(0.23 \pm 0.01)$, indicating that the inhomogeneously broadened
absorption profile of Pr$^{3+}$ also contributes to partially
filter the noise \cite{Rielander2014}. Nonetheless, for this set
of measurements, we were able to achieve higher storage
efficiency, i.e.  $\eta_{SW} = (5.3 \pm 0.5)\,\%$, leading to a
limited increase of the $\mu_1$ to about 4.

For applications in QIS, it is crucial that the
optical memory preserves the coherence of the stored qubits. We
take advantage of the intrinsic temporal multimodality of the AFC
protocol to demonstrate the phase preservation in the spin-wave
storage of time-bin qubits. This type of encoding is widely used
in quantum communication as it is robust against decoherence in
optical fibers \cite{Marcikic2003}. The time-bin qubits are
expressed as $|\psi_{in}\rangle=c_1|e\rangle+c_2e^{i\Delta
\alpha}|l\rangle$, where $|e\rangle$ ($|l\rangle$) represents a
qubit in the early (late) time-bin, $\Delta \alpha$ is their
relative phase, and $c_1^2+c_2^2=1$. In order to store time-bin
qubits, the duration of the input pulses is reduced (from $430$ to
$260\,\mathrm{ns}$), leading to a reduction of $\eta_{SW}$ to
about $2.2\,\%$ and to an increase of the $\mu_1$ up to $\mu_{1p}
= 0.11 \pm 0.01$. We start by evaluating the fidelity of the
states $|e\rangle$ and $|l\rangle$, located at the poles of the
Bloch sphere, $F_{e}$ and $F_{l}$, by storing only the early and
the late qubits. We obtain average fidelity values for the poles
ranging from $F_{el}=85\,\%$ to $98\, \%$ for photon number per
qubit, $\mu_q$, going from 0.6 to 5.9 \cite{SupplMat}.\\
\begin{figure}[h]
\centering\includegraphics[width=1\columnwidth]{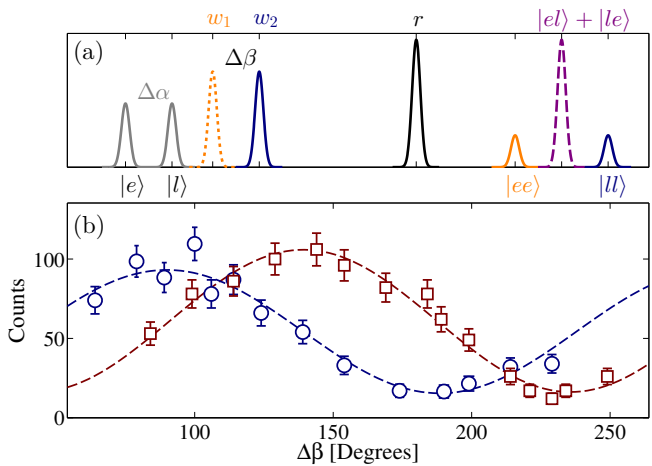}
\caption{(a) Pulse sequence to measure the
time-bin qubit coherence. We apply two partial write pulses with a
relative phase $\Delta \beta$ in order to split each pulse into
two temporally separated echoes. If the delay between the time-bins
 $|e\rangle$ and $|l\rangle$ is equal to the time
difference between the two write pulses, we can overlap the late
echo of the early bin, $|el\rangle$, with the early echo of the
late bin, $|le\rangle$. The output of the memory (occurring after
the single read pulse $r$) has three time-bins, $\{|ee\rangle,
|el\rangle + |le\rangle,|ll\rangle\}$. An interference will occur
in the central time-bin if the coherence is preserved. The input
pulses, located at $|e\rangle$ and $|l\rangle$ have a relative
phase difference of $\Delta\alpha$. (b) Interference fringes
obtained integrating over the central output time-bin (in this
case $\Delta t_d = 0.5 \,\mathrm{\mu s}$) as a function of the
relative phase difference $\Delta\beta$ for $\mu_q = 1.5$. Circles: $\Delta\alpha = 90\,^{\circ}$, $V =
(71.6 \pm 6.8) \,\% $; Squares: $\Delta\alpha = 135\,^{\circ}$, $V =
(73.4 \pm 3.5) \,\% $.}
\label{time-bin}
\end{figure}
We then store superposition states located on the equator of the
Bloch sphere. We use the memory itself to analyze the retrieved
qubits \cite{Timoney2013} applying two partial write pulses as depicted in
Fig. \ref{time-bin}(a). This method provides a convenient way of
analyzing time-bin qubits, but has the drawback of reducing the
storage efficiency. As a matter of fact, in order to insert two
write pulses, their duration needs to be reduced, which decreases
their efficiency. Fig. \ref{time-bin}(b) reports examples of
interference fringes for $\mu_q = 1.5$.
From sinusoidal fits, we obtain a raw mean visibility of $V_{+-} =
(72.5 \pm 1.3) \,\% $. The fidelity of the process is calculated
from the visibility, as $F = (1+V)/2$ \cite{Marcikic2003}. Finally,
we obtain a total conditional fidelity per retrieved qubit $F_T$ = $\frac{1}{3}$$
F_{el}$ + $\frac{2}{3}$$ F_{+-}$, where $F_{el}$ ($F_{+-}$) is the
average fidelity over the poles (equator) basis. The obtained
values are reported in Fig. \ref{FvsU} for different $\mu_q$. We
observe that the fidelity decreases with $\mu_q$. To explain this
behavior, we fit our data with a simple model taking into account
the decrease of SNR with $\mu_q$ and the reduced efficiency due to
the double write protocol \cite{SupplMat}. The good
agreement between the simple model and the data provides evidence
that the decrease of fidelity is only due to the noise created by
the control pulses, and not to a loss of coherence.

In order to infer the quantum nature of our memory, the total
fidelity is compared with the highest fidelity achievable with a measure-and-prepare approach (solid curve in Fig. \ref{FvsU}) taking into account the Poissonian statistics of the
input states and the finite memory efficiency ($2.2\,\%$)
\cite{Specht2011,Gundogan2012,Nicolas2014}. Using this criteria, the experimental data
are higher than the limit for a classical memory by more than one standard
deviation for most $\mu_q$ investigated. With the raw data, the
memory is in the quantum regime for $\mu_q
> 0.96$. When correcting for the loss of efficiency in the
analysis (which could be achieved by analyzing the qubits with an
external interferometer), the model predicts that the quantum
regime would be reached for $\mu_q > 0.25$ (see \cite{SupplMat}).

The very low noise probability and the ability to obtain $\mu_1 \ll
1$ opens prospects for the spin-wave storage of single-photons, as
required for many applications in QIS. In
that case, the probability to have a photon before the memory (i.e
including all optical loss between source and memory) needs to be
higher than $\mu_1$ to enter the quantum regime. In the current
experiment, the storage efficiency is limited by the available
comb optical depth in our 3 mm long crystal, by the limited
transfer efficiency ($\eta_T^2 =(67 \pm 4)\,\%$) and by technical issues
(cryostat vibrations, laser linewidth, see \cite{SupplMat}). Note that
much higher efficiencies (for storage in the excited state) have been
obtained in \PrYSO using longer crystal (69 $\%$)
\cite{Hedges2010} or impedance matched cavities
(58 $\%$) \cite{Sabooni2013}. Longer storage times can also be achieved with dynamical decoupling techniques to
counteract decoherence in the spin state \cite{Longdell2005,Heinze2013,Pascual-Winter2012,Lovric2013}.

\begin{figure}[H]
\centering\includegraphics[width=0.9\columnwidth]{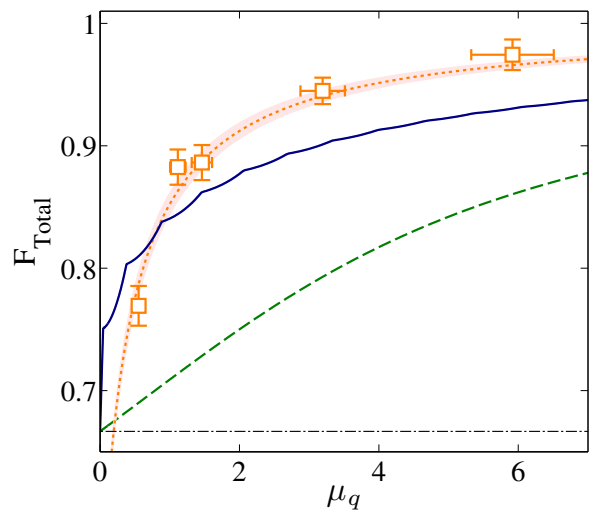}
\caption{Total fidelity vs input photon number per qubit, $\mu_q$. The
light orange squares are the data points
 with an error bar of 1 standard deviation. The orange dotted line is a fit to the data points
using Eq. (5) in \cite{SupplMat}, with the
corresponding shaded area being the 1 standard deviation of the
error in this fit. The solid blue (dashed green) line is the
classical limit obtained by a measure and prepare strategy for our
memory efficiency of $\eta_{SW} = 2.2 \, \%$ ($\eta_{SW} = 100 \,
\%$) when testing the memory with weak coherent states
\cite{Gundogan2012}. The dash-dotted line is the classical limit
for testing the memory with a single-photon Fock state $(F =
2/3)$.} 
\label{FvsU}
\end{figure}

In conclusion, we demonstrated the spin-wave storage and on-demand
retrieval of weak coherent states at the single-photon level in a
solid state memory based on a \PrYSO crystal. This is the first demonstration that solid state spin-wave optical memories can operate in the quantum regime, overcoming a strong limitation for AFC QMs. We achieved a SNR
higher than 10 for single-photon level input pulses, high enough to enable the storage of single photons.
Finally, we confirmed the quantum nature of our
memory by storing time-bin qubits encoded in weak coherent states
and demonstrating conditional fidelities for the retrieved qubits higher than what is possible with classical
memories.  Our device thus represents the first spin-wave memory for photonic time-bin qubits. These results open the door for long-lived storage and
on-demand readout of non-classical light states in solid state
devices and represent an important step in view of using solid
state quantum memories in scalable quantum architectures.

\textbf{Acknowledgements.} We acknowledge
financial support by the ERC starting grant QuLIMA, by the Spanish
Ministry of Economy and Competitiveness (MINECO) and the Fondo
Europeo de Desarrollo Regional (FEDER) through grant
FIS2012-37569, by the European project CHIST-ERA QScale and by the
People Programme (Marie Curie Actions) of the EU FP7 under REA
Grant Agreement No. 287252.

\hrulefill
\newpage

\hrulefill

\appendix

\section{Supplemental Material}

\maketitle

In the supplemental material we report details on the atomic frequency comb preparation (section I.), the sources of noise (section II.), the chosen strategy to suppress the technical noise (section III.), the characterization of the filter cristal (section IV.), and, finally, we describe a theoretical model for the fidelity of the spin-wave storage of time-bin qubits (section V.).

\section{Atomic Frequency Comb Preparation}\label{Atomic Frequency Comb Preparation}

\begin{figure}[h]
\centering\includegraphics[width=0.85\columnwidth]{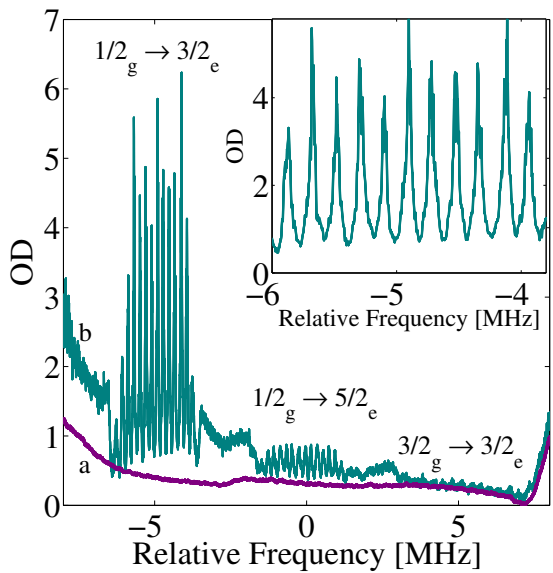}
\caption{Example of transparency window (curve a) and comb trace prepared with $\Delta=200 \,\mathrm{kHz}$ (curve b). The inset shows a magnification of the comb corresponding to the ${1}/{2}_\textrm{g}-{3}/{2}_\textrm{e}$ transition. }
\label{comb}
\end{figure}

To prepare the atomic frequency comb (AFC) we follow the approach already described in \cite{Timoney2013,Maring2014}. The frequency of the strong beam is firstly scanned by 14 MHz in 100 ms to create a wide transparency window (also referred to as pit) in the Pr$^{3+}$ absorption line by spectral hole burning (curve a in Fig. \ref{comb}). Afterwards a $2\,\mathrm{MHz}$-wide sweep is operated outside the pit to pump the atoms back to the ${1}/{2}_\textrm{g}$ state. This creates a $3.5\,\mathrm{MHz}$-wide absorbing feature corresponding to the ${1}/{2}_\textrm{g}-{3}/{2}_\textrm{e}$ transition inside the pit, but also populates the ${3}/{2}_\textrm{g}$ state which in principle must be empty for later transfer of the input field (see Fig. 1(b) in the main text). Thus, a clean pulse with duration of 50 ms is applied on ${3}/{2}_\textrm{g}-{3}/{2}_\textrm{e}$ transition to empty this ground state. The clean pulse also contributes to suppress the transitions of different classes of Pr$^{3+}$ ions which might be addressed by the preparation pulses \cite{Guillot-Noel2007}. Then, a sequence of single-frequency hole-burning pulses is applied on the single class absorption feature resonant with the ${1}/{2}_\textrm{g}-{3}/{2}_\textrm{e}$ transition, each time changing the frequency by a fixed amount, $\Delta$. This sequence burns periodically spaced holes in the absorbing feature corresponding to the ${1}/{2}_\textrm{g}-{3}/{2}_\textrm{e}$ transition and anti-holes at the frequency of the ${3}/{2}_\textrm{g}-{3}/{2}_\textrm{e}$ transition, so a short burst of clean pulses is applied to maintain the ${3}/{2}_\textrm{g}$ state empty. For the AFC preparation it is crucial to have a third ground state (${5}/{2}_\textrm{g}$) to use as auxiliary state where to store the excess of population resulting from the optical pumping. An example of the resulting comb structure for $\Delta=200 \,\mathrm{kHz}$ is shown in Fig. \ref{comb}, curve b. With the described procedure we are able to tailor absorption peaks as narrow as $(43 \pm 3)\,\mathrm{kHz}$, with a comb optical depth $d= 4.5 \pm 0.1$ and background optical depth $d_0= 0.75 \pm 0.04$. It is worth noting that the limited dynamical range of the detector might fix a higher bound for the detected peak amplitude and thus lead to an overestimated full width at half maximum.

When the memory is characterized  with classical pulses, we
find typical efficiency values of $\eta_{AFC} = (10.3 \pm 1.4) \%$
and $\eta_{SW} = (4.4 \pm 0.1) \%$. The prediction for Gaussian
AFC peaks is $\eta_{AFC} =
\tilde{d}^2e^{-\tilde{d}}e^{-7/F^2}e^{-d_0}$, where $\tilde{d} =
d/F$ is the effective comb optical depth \cite{Afzelius2009}. The
measured value of $\eta_{AFC}$ is compatible with the prediction,
$\eta_{AFC,th} = (12.2 \pm 1.6)\,\%$,  for the measured values of
$d = 4.5 \pm 0.1$, $F = 4.7 \pm 0.4$ and $d_0 = 0.75 \pm0.04$. For the same optical depth of
our comb but assuming $d_0 = 0$, the model predicts a maximal
efficiency of $\eta_{AFC,th} = 26.5\,\%$ for a finesse of $F = 4$. 
Note that an even higher efficiency, $\eta_{AFC,th} = 34.8\,\%$, is expected for square shaped comb peak and a finesse of $F = 3.2$ (\cite{Bonarota2010}). In our case, the
achievable contrast is limited by technical reasons, such as
vibrations of the cryo-cooler and finite linewidth of the laser.
When measured in photon counting mode, the efficiency is further
reduced by the coincidence window taken (containing $\sim 80\,\%$
of the total counts), by loss in the narrow-band filter ($\sim
10\,\%$) and by instabilities due to long integration times ($\sim
10\,\%$ loss).

\section{Sources of noise}\label{Noise sources}
\begin{figure}[h]
\centering\includegraphics[width=1\columnwidth]{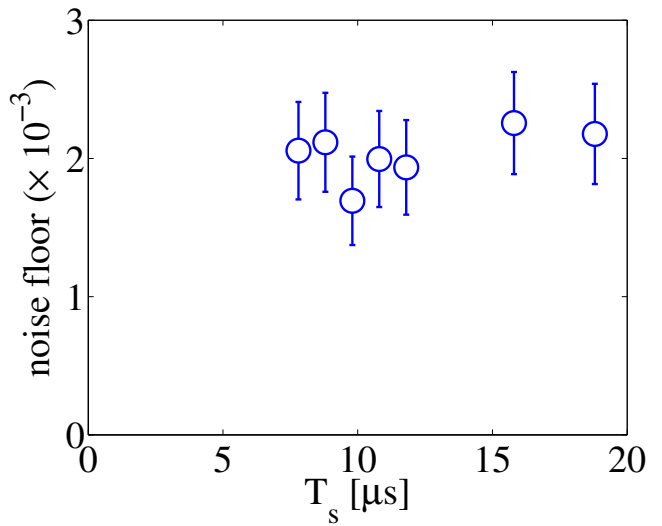}
\caption{Noise floor in the echo window as a function of the time between the two control pulses, $T_s$. The error bars are calculated taking into account Poissonian statistics.}
\label{noisevsTs}
\end{figure}

The full AFC protocol for light storage implies the use of strong transfer pulses which
are sources of different kind of noise. First of all there might be spatial leakage from the preparation/control mode into the input mode, essentially due to reflections and scatting from the optical surfaces.
Moreover, the control pulses might interact with the residual population left in the ${3}/{2}_\textrm{g}$ state, due to imperfect cleaning (see section \ref{Atomic Frequency Comb Preparation}). This interaction can give rise to coherent (e.g. free-induction decay) or incoherent emission. 
Another potential source of noise in this system may arise from four-wave mixing, where the first control pulse creates spin-waves via an off-resonant Raman interaction and the second control pulse reads them out as noisy photons in the same mode as the memory output. This has been identified as the main source of noise in Raman type quantum memories using atomic vapors \cite{Michelberger2015}. However, this type of noise does not seem to be, at this stage, an important source of noise in our spin-wave memory. This is evidenced by the following noise measurements.
First, we have measured that the two control pulses contribute equally to the noise in the echo window. This can be explained by the fact that they interact with the same number of ions giving two equally sized incoherent fluoresecence fields (given the time scale of the storage time compared to the excited state lifetime $T_1$ = 164 $\mu s$ \cite{McAuslan2009}). If four-wave mixing was not negligible, the noise would be suppressed by more than a factor 2 by switching off one of the two control pulses. Furthermore, the noise arising from four-wave mixing would decay for increasing storage times with the same behavior of the spin-wave echo, i.e. following the inhomogeneous spin broadening (see Fig. 2(c) of the main text). As a matter of fact, the noise floor in our case is constant within error over the whole range of storage times investigated (see Fig. \ref{noisevsTs}).

Several reasons could explain this difference. First, the non-collective emission from the $^{1}$D$_{2}$  excited state mainly happens through intermediate crystal-field manifold and involves the coupling with the phonons of the YSO matrix. As a matter of fact, the branching ratio of the spontaneous emission through the $ ^{1}$D$_{2}\rightarrow ^{3}$H$_{4}$ transition is very low ($\approx 3\,\%$ ) \cite{McAuslan2009}. Furthermore, among the transitions within the $^{3}$H$_{4} \rightarrow ^{1}$D$_{2}$ that could be off-resonantly excited by the first control pulse, many of them are characterized by relative oscillator strengths one order of magnitude lower than that employed to transfer the excitation to the ground state. During the AFC preparation, most of the ions are collected in the auxiliary ${5}/{2}_\textrm{g}$ state but, due to the relative oscillator strength of the transitions starting from it, they can only be involved in the almost closed cycle transition ${5}/{2}_\textrm{g} \leftrightarrow {5}/{2}_\textrm{e}$ \cite{Nilsson2004}.
Those characterized by a non-negligible probability for off-resonant excitation either are suppressed by the state cleaning that we operate in the memory preparation (it is the case of the transitions starting from the  ${3}/{2}_\textrm{g}$  state) or are tailored in the comb structure (transitions having the ${1}/{2}_\textrm{g}$ as initial state). In the latter case, the second control pulse would give rise to a coherent emission in a well separated temporal mode with respect of the spin-wave echo. 
\section{Filtering}
To achieve a low noise floor, $(2.0 \pm 0.3) \times 10^{-3}$, we use
spatial, temporal and spectral filtering. As for the spatial
filtering, the control and input beams are sent to the memory with
a small angle, leading to an extinction of $10^{-5}$. Temporal
filtering is achieved thanks to a temporal gate implemented with
two AOMs. It allows us to block the leakage of the strong control
beams in the signal mode. This is important not to blind the
single-photon detectors, but also to avoid burning a spectral hole
in the crystal filter. A diffraction grating is employed to decrease the 
noise originated from incoherent fluorescence. The narrow-band spectral filter is prepared
simultaneously with the AFC. The frequency of the filter mode
(shown in dashed red line in Fig. 1(a)) is scanned by
$1.2\,\mathrm{MHz}$ around the input light frequency for 100 ms
which results in a transparency window of around $2\,\mathrm{MHz}$
due to the power broadening effect. In order to temporally
discriminate the stored and retrieved weak pulses from the portion
of the FID leaking in the echo mode and happening at short time
scales (a few $\mathrm{\mu s}$), the optimization of the excited
state storage efficiency at long storage times is crucial. By
carefully optimizing the storage times, in both the excited and
ground states, and the pulse durations a compromise can be reached
between the noise level and a reasonable retrieval efficiency.

\section{Filter crystal characterization}

We provide in this section details about the noise suppression realized by the filter crystal. The main advantage of using a second crystal is that it provides a dynamical filter, whose center frequency and bandwidth can be easily tuned, as it is obtained by spectral hole burning. 

\begin{figure}[h]
\centering\includegraphics[width=1\columnwidth]{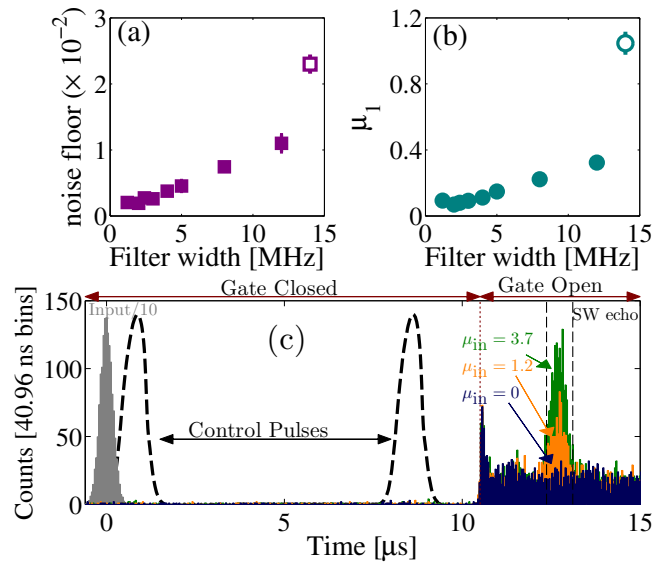}
\caption{Noise floor in the echo window (a) and $\mu_1$ (b) value as a function of the spectral hole width in the filter crystal. Filled symbols: hole centered at the input pulse frequency; open symbols: transparency window coincident with that prepared in the memory crystal (see curve b in Fig. \ref{comb}). (c) Time histogram of the retrieved photons for different input photon numbers, $\mu_{in}$, and a filter width of $14 \,\mathrm{MHz}$. The input ($\mu_{in} = 0.9$) and the control pulses (measured with a reference photodetector) are also displayed.}
\label{filter}
\end{figure}
Fig. \ref{filter}(a) represents the noise floor (noise counts per pulse at the memory crystal) as evaluated from noise measurements where we burn in the filter crystal spectral holes of different width centered about the input pulse frequency (solid symbols). The noise in the echo window increases by one order of magnitude when the filter width increases from 0.8 to 12 MHz. The open symbol refers to the extreme case of a 14 MHz-wide transparency window coincident with the one prepared in the memory crystal to host the AFC (see curve a in Fig. \ref{comb}). Here the noise floor is further doubled because the transparency window allows the control pulse frequency to pass by and reach the detector. The increase in the noise affects consequently the $\mu_1$ value as shown in Fig. \ref{filter}(b). Nevertheless, this value also depends on the echo counts, thus the narrowest filter width does not necessarily correspond to the best case. As a matter of fact, the best $\mu_1$ value is measured for a filter width which allow the major portion of the echo signal to pass by but suppresses efficiently the noise at the control pulse frequency, i.e. $2 \,\mathrm{MHz}$. The value of $\mu_1$ also depends on the echo efficiency and is thus more affected by power fluctuations than the noise floor. We believe this is the reason why its increase in the case of totally open transparency window is more pronounced than that of the noise floor (compare filled and open symbols in panel (a) and (b) of Fig. \ref{filter}). Panel (c) of Fig. \ref{filter} shows the time histogram of the retrieved photons for different input photon numbers in the case of a  $14 \,\mathrm{MHz}$-wide pit prepared in the filter crystal. Finally we evaluate the extinction ratio of our $2 \,\mathrm{MHz}$-wide spectral filter by switching off the AFC and sending input pulses through the transparency window in the memory crystal firstly resonant with the filter center, then at the control pulse frequency. By comparing the total counts reaching the detector in the two cases, we estimate the extinction ratio to be about 750. 

\section{Modeling the Fidelity}

Fig. 4 from the main document shows the measured total fidelity
as a function of the mean photon number per qubit, $\mu_q$, with a corresponding fit
using the model we develop in this section. The aim of the model
is to show that the reduction of the fidelity as the photons per
qubit are reduced is entirely due to the noise of the memory and
not due to a loss of coherence caused by e.g. phase noise in the
laser. The goal of our model is to express the total fidelity as
a function of the same $\mu_{1}$ parameter.

\begin{figure}[ht]
\centering
\subfigure[]{
\centering\includegraphics[width=0.4\textwidth]{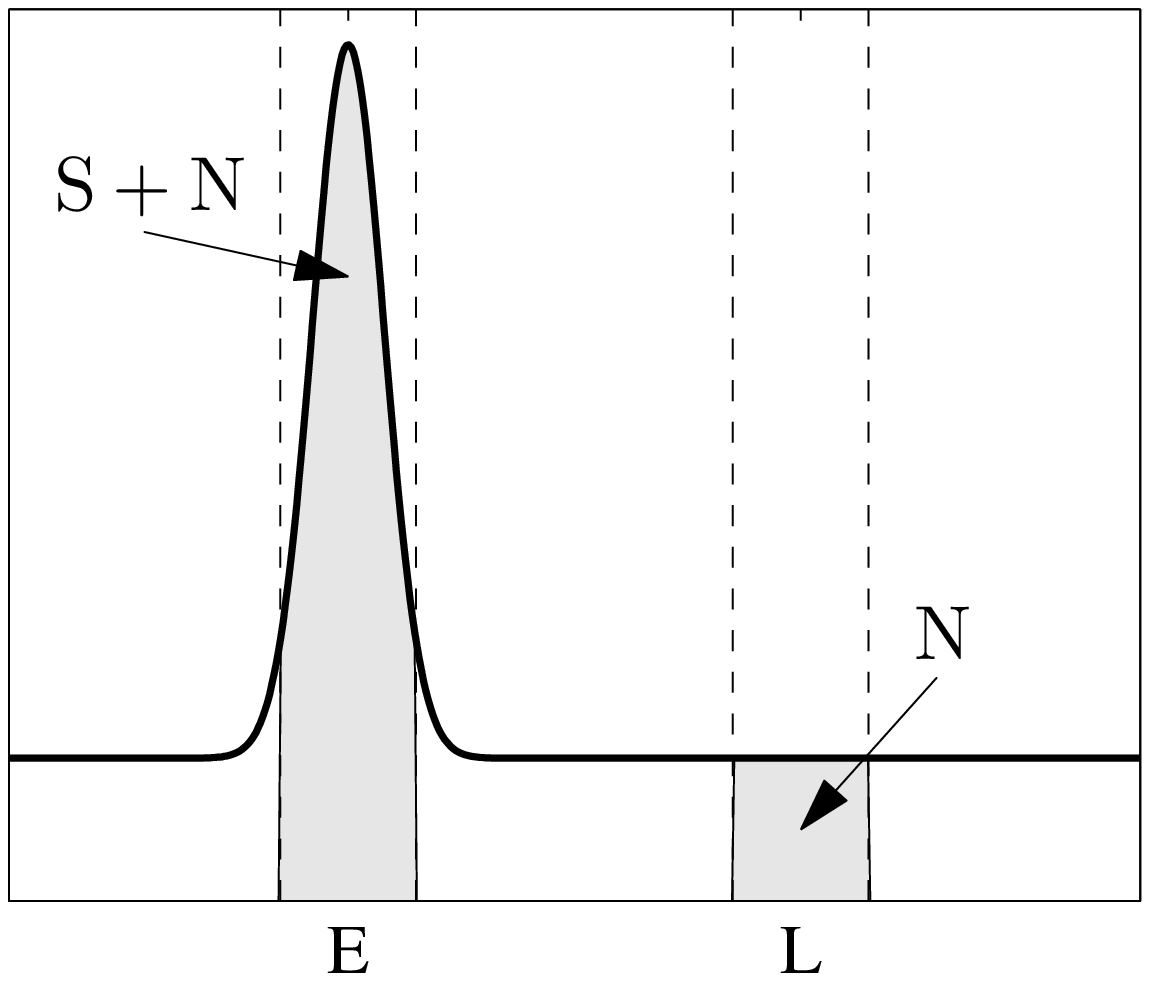}
    \label{fig:subfig1}
}
\subfigure[]{
\centering\includegraphics[width=0.4\textwidth]{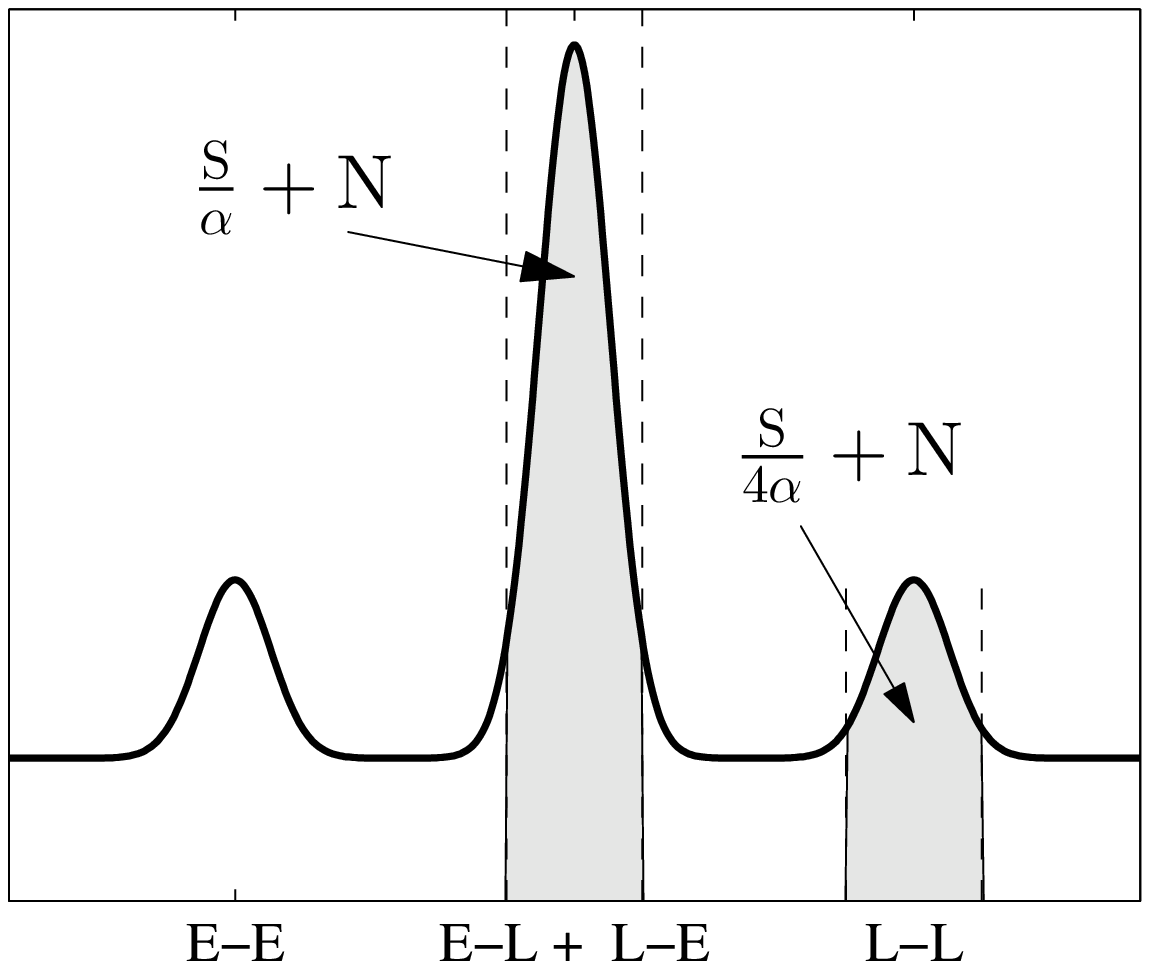}
    \label{fig:subfig2}
}
\caption{\subref{fig:subfig1} Schematic example of the memory output for storage of an $|e\rangle$ qubit. \subref{fig:subfig2} Schematic example of the memory output for storage of an $|e\rangle + |l\rangle$ where the relative phases are such that the maximum interference is observed.}
\label{fig:1}
\end{figure}

Firstly, let us look at the output of the memory for the two
differing measurements, namely the measurement of the $|e\rangle$
or $|l\rangle$ qubit (i.e. the poles of the Bloch sphere) and the
measurement of the coherence between $|e\rangle$ and $|l\rangle$
(i.e the equators of the Bloch sphere). The former (latter) are
schematically depicted in Fig. \ref{fig:subfig1}
(\ref{fig:subfig2}).

For a given integration width, we sum over both E and L for
the poles measurement, while for the coherence measurement we
integrate over the window labelled E-L$+$L-E (see figure
\ref{fig:subfig2}).

Consider the example depicted in Fig. \ref{fig:subfig1}, where
the $|e\rangle$ qubit has been stored and then recalled. We label
the integrated counts in the window about E as $\mathrm{S} +
\mathrm{N}$ and the integrated counts about the window L as
$\mathrm{N}$, where $\mathrm{S}$ is the signal (not including the
noise) and $\mathrm{N}$ is the noise. Then, the fidelity is the
ratio of the counts in E to the total counts over E and L, i.e.
\begin{equation}
\F{e} = \frac{\mathrm{S}+\mathrm{N}}{\mathrm{S}+\mathrm{N} \,\,\,\,+\,\,\,\,\mathrm{N}} = \frac{\mathrm{S}/\mathrm{N} + 1}{\mathrm{S}/\mathrm{N} + 2} = \frac{\SNR + 1}{\SNR + 2}.
\end{equation}
where $\SNR$ is the signal to noise ratio. Provided the efficiency
and noise are the same for storing the $|l\rangle$ qubit, $\F{l} =
\F{e} = \F{el}$. Finally, we can express the above fidelity as a
function of the $\mu_1$ of this measurement (see main text for
definition) by noting that $\SNR = \mu_q/\mu_{1p}$, where
$\mu_{1p}$ is the $\mu_1$ characteristic of the poles measurement.
The expression for the fidelity becomes
\begin{equation}
\F{el} = \frac{\mu_q + \mu_{1p}}{\mu_q + 2\mu_{1p}}.
\end{equation}
and we measure $\mu_{1p} = 0.11 \pm 0.01$.

Now let us consider Fig. \ref{fig:subfig2}, which depicts an
arbitrary qubit $|e\rangle + |l\rangle$ being recalled, and the
relative phases are such that the maximum interference is
observed. Firstly, the integrated counts corresponding to the
window at the time labelled E-E are $\mathrm{S}/4\alpha +
\mathrm{N}$ where the signal $\mathrm{S}$ is the same as defined
earlier. Here, the signal $\mathrm{S}$ is reduced by a factor of
4, this is because the number of photons per qubit are distributed
equally within the input time-bins (factor of 2) and the echo is
distributed equally between the output time-bins (factor of 2).
The factor $\alpha$ reduction in the signal is because we use
different bandwidth pulses for the double write process which
gives a reduced efficiency in the echo. The amount of signal in
the window E-L + L-E is 4 times that of the signal in the L-L
window provided the interference is fully coherent (note the same is true
for the E-E window).The resulting counts are then
$\mathrm{S}/\alpha + \mathrm{N}$. It is worth noting that if we
instead used an unbalanced Mach Zehnder interferometer, we would
have no reduction in the signal and $\alpha = 1$.

The fidelity of the coherence measurement is obtained from the
visibility of the interference fringe by $\F{}~=~(1~+~V)/2$ \cite{Marcikic2003} where
the visibility $V$ is defined as
\begin{align}
V &= \,\frac{\mathrm{max} - \mathrm{min}}{\mathrm{max} + \mathrm{min}} = \,\frac{\mathrm{S}/\alpha + \mathrm{N} \,\,\,\, - \,\,\,\, \mathrm{N}}{\mathrm{S}/\alpha + \mathrm{N} \,\,\,\, + \,\,\,\, \mathrm{N}} \nonumber\\
&= \frac{\mathrm{S}}{\mathrm{S}+ 2\mathrm{N}\alpha} = \frac{\SNR}{\SNR + 2\alpha}
\end{align}
where max (min) refers to the maximum (minimum) integrated counts
of the fringe, the max being $\mathrm{S}/\alpha +\mathrm{N}$ and min being
$\mathrm{N}$ . Note
that we have $V$ in terms of the $\SNR$ of the poles measurement,
so the resulting fidelity can be written in terms of $\mu_{1p}$,
i.e.
\begin{equation}
\F{+-} = \frac{1}{2} + \frac{1}{2} \,\frac{\mu_q}{\mu_q + 2\alpha\mu_{1p}},
\end{equation}
where the subscript $+ (-)$ refers to the qubit $|e\rangle \pm |l\rangle$ ($|e\rangle \pm i |l\rangle$), and we assume $\F+ = \F- = \F{+-}$. We measure $\alpha$ by comparing the $\mu_1$ in the L-L window of figure \ref{fig:subfig2} to $\mu_{1p}$ and we get $\alpha = 2.5 \pm 0.6$.

\begin{figure}[h]
\centering\includegraphics[width=1\columnwidth]{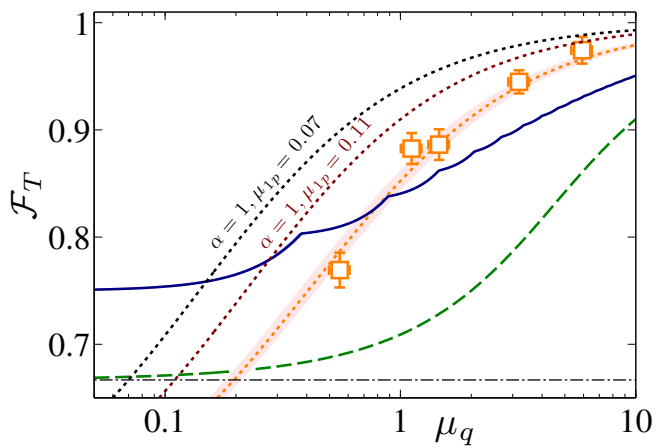}
\caption{Total fidelity vs input photon number per qubit. The light orange squares are the data points with an error bar of 1 standard deviation. The light orange dotted line is a fit to the data points using eq. \ref{eq:Ftotal} with $\mu_{1p}$ fixed at $0.11$ and $\alpha = 2.5 \pm 0.3$. The corresponding shaded area is the 1 standard deviation of the error in this fit. The solid blue (dashed green) line is the classical limit obtained by a measure and prepare strategy and a memory efficiency of $\eta_{\textrm{3LE}} = 2.2 \, \%$ ($\eta_{\textrm{3LE}} = 100 \, \%$) when testing the memory with weak-coherent states \cite{Specht2011,Gundogan2012}. We see that for $\mu_q > 0.96$ the fidelity is above the classical benchmark. We show also the case of $\alpha = 1$ and $\mu_{1p} = 0.11$, i.e. the case where our detection of the coherence is replaced by an unbalanced Mach Zehnder interferometer for example. Furthermore, we show the case of $\alpha = 1$ and $\mu_{1p} = 0.07$, i.e. the $\mu_1$ measured for the optimised pulse bandwidth case. The dash-dotted line is the classical limit for testing the memory with a single-photon Fock state $(\F{} = 2/3)$.}\label{fig:FvsU}
\end{figure}

Now, the total fidelity is given by $\F{T} = \frac{1}{3}\F{el} + \frac{2}{3}\F{+-}$, which leads to
\begin{equation}
\F{T} = \frac{1}{3}\left[\frac{\mu_q + \mu_{1p}}{\mu_q + 2\mu_{1p}}\right] + \frac{1}{3} \left[1+ \,\frac{\mu_q}{\mu_q + 2\alpha\mu_{1p}}\right], \label{eq:Ftotal}
\end{equation}
concluding our simple model for the fidelity. Taking the measured $\mu_{1p}$, we fit the data in Fig. \ref{fig:FvsU} with equation \ref{eq:Ftotal} having $\alpha$ as a free parameter. We get $\alpha = 2.5 \pm 0.3$ which agrees excellently with the measured value stated earlier. We therefore can conclude that the reduction of the fidelity is due to the noise on the output of the memory and not phase noise. For $\mu_q > 0.96$ the fidelity surpasses the classical bound, remaining in the quantum regime. Also in Fig. \ref{fig:FvsU} we plot the case for $\alpha = 1$ for $\mu_{1p} = 0.11$ and $\mu_{1p} = 0.07$ (the case shown in Fig. 2 of the main text). The fidelity exceeds the classical bound for photon numbers much less that 1 for these cases.

\begin{figure}[htbp]
\begin{center}
\centering\includegraphics[width=1\columnwidth]{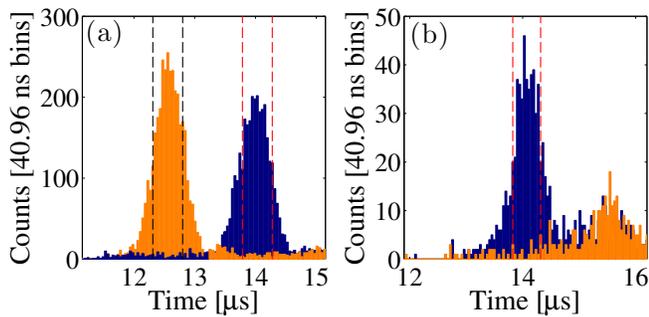}
\caption{(a) An example of the detected counts at the output of the memory for the early (light orange histogram) and late (dark blue histogram) input qubits with $\mu_q = 3$. The vertical dashed lines correspond to the $0.5\,\mu$s integration windows for the calculation of the fidelity. (b) An example of detected counts at the output of the memory for an input qubit $|e\rangle + |l\rangle$ with $\mu_q = 5.9$. The dark blue (light orange) histogram shows the case when the relative phase of the double-write process is such that the interference is maximized (minimized) in the integration window (indicated by the vertical dashed line). The pulse feature on the right corresponds to the L-L window of Fig. \ref{fig:subfig2} while the E-E window is not detected here as the temporal gate is closed to negate the effects of the control pulses.}
\label{fig:PolesAndCoh}
\end{center}
\end{figure}

For completeness, we include table \ref{tab:fidelity} which shows all the measured fidelities for every $\mu_q$ tested. Also, we include examples of the output of the memory for early and late inputs (Fig. \ref{fig:PolesAndCoh}(a)) and for the input $|e\rangle + |l\rangle$ where the phase of the double-write process is adjusted to show maximum and minimum interference in the E-L + L-E window.

\begin{table}[H]
\vspace{5mm}
\centering
\begin{tabular}{ c  c  c  c c}
\hline
\hspace{1mm} $\mu_q$\hspace{1mm} & $\F{el}$ & $\F{+-}$ & $\F{T}$  & $\F{C}$ \\
\hline
\hline
$5.9$ & \hspace{0.4mm} $(97.9 \pm 1.5)\%$ \hspace{0.4mm} & \hspace{0.4mm} $(97.2 \pm 1.7)\%$ & \hspace{0.4mm} $(97.4 \pm 1.2)\%$ \hspace{0.4mm} & $(93.0\pm 0.1)\%$ \hspace{0.4mm} \\
$3.2$ & \hspace{0.4mm} $(96.9 \pm 2.2)\%$ \hspace{0.4mm} & \hspace{0.4mm} $(93.3 \pm 1.2)\%$ & \hspace{0.4mm} $(94.5 \pm 1.1)\%$ \hspace{0.4mm} & $(90.1\pm 0.1)\%$ \hspace{0.4mm}\\
$1.5$ & \hspace{0.4mm} $(93.5 \pm 2.9)\%$ \hspace{0.4mm} & \hspace{0.4mm} $(86.2 \pm 1.6)\%$ & \hspace{0.4mm} $(88.6 \pm 1.4)\%$ \hspace{0.4mm} & $(86.2\pm 0.1)\%$ \hspace{0.4mm}\\
$1.1$ & \hspace{0.4mm} $(93.3 \pm 3.1)\%$ \hspace{0.4mm} & \hspace{0.4mm} $(85.8 \pm 1.5)\%$ & \hspace{0.4mm} $(88.3 \pm 1.4)\%$ \hspace{0.4mm}&  $(84.4\pm 0.1)\%$ \hspace{0.4mm} \\
$0.6$ & \hspace{0.4mm} $(84.9 \pm 4.2)\%$ \hspace{0.4mm} & \hspace{0.4mm} $(72.9 \pm 1.3)\%$ & \hspace{0.4mm} $(76.9 \pm 1.6)\%$ \hspace{0.4mm} & $(81.0 \pm 0.1)\%$ \hspace{0.4mm}\\
\hline
\end{tabular}
\caption{Fidelities $\mathcal{F}_{el}$, $\mathcal{F}_{+-}$ and $\mathcal{F}_{T}$ for each $\mu_q$ tested. $\F{C}$ is the classical benchmark for testing the device with weak coherent states which depends on the efficiency of the memory and the input photon number used \cite{Specht2011,Gundogan2012}. The stated errors are one standard deviation.\label{tab:fidelity}}
\end{table}

\bibliographystyle{prsty}

\end{document}